\documentclass[preprint,showkeys,aps,prb]{revtex4}
\usepackage[dvips]{graphicx}
\begin{document}

\title{                                                                                                                   
Thermodynamic Entropy from Sadi Carnot's Cycle using \\
Gauss' and Doll's-Tensor  Molecular Dynamics
}


\author{
William Graham Hoover with Carol Griswold Hoover \\
Ruby Valley Research Institute                   \\
Highway Contract 60, Box 601                     \\
Ruby Valley, Nevada 89833                        \\
}

\date{\today}


\vspace{0.1cm}

\begin{abstract}
Carnot's four-part ideal-gas cycle includes both isothermal and adiabatic expansions
and compressions. Analyzing this cycle provides the fundamental basis for statistical
thermodynamics. We explore the cycle here from a pedagogical view in order to promote
understanding of the macroscopic thermodynamic entropy, the state function associated
with thermal energy changes.  From the alternative microscopic viewpoint the Hamiltonian
${\cal H}(q,p)$ is the energy and entropy is the (logarithm of the) phase-space volume 
$\Omega$ associated with a macroscopic state. We apply two novel forms of Hamiltonian
mechanics to Carnot's Cycle: [1] Gauss' isokinetic mechanics for the isothermal segments
and [2] Doll's Tensor mechanics for the isentropic adiabatic segments. We explore the
equivalence of the microscopic and macroscopic views of Carnot's cycle for simple fluids
here, beginning with the ideal Knudsen gas and extending the analysis to a prototypical
simple fluid.
\end{abstract}

\maketitle

\section{From Phase Space to Equilibrium Thermodynamics}
Boltzmann and Gibbs showed that macroscopic thermodynamics, that set of mechanical 
and thermal relations linking heat, work, temperature, energy, pressure, and volume, 
is a straightforward consequence of atomistic models of matter. We strengthen that
connection here using two relatively recent versions of Hamiltonian mechanics. Gauss'
Principle of Least Constraint [ applied to the kinetic energy ] is the first. It
provides a deterministic and time-reversible basis in Hamiltonian mechanics for
{\it isothermal} simulations\cite{b1}. Doll's Tensor\cite{b2} is the second. It
incorporates the macroscopic volumetric strain rate, $(\dot V/V) = (\dot x/x) + (\dot y/y) =
2\dot \epsilon$, necessary to simulating mechanical work with {\it adiabatic}
molecular dynamics. These two twentieth-century developments strengthen and illuminate
the nineteenth-century connection between microscopic atomistic particle models and
macroscopic thermodynamics based instead on empirical constitutive equations of state.
Adding the microscopic kinetic-theory notions of ideal-gas thermometry to macroscopic
thermodynamic notions of temperature and entropy through Gauss-Principle isothermal and
Doll's-Tensor adiabatic deformations provides a straightforward atomistic derivation of
macroscopic thermodynamics!

For simple fluids macroscopic equilibrium thermodynamics describes two different
kinds of energy changes, $dE$: [1] heat taken in, $dQ$, and [2] mechanical work
done, $dW$. For a ``reversible'' series of equilibrium states the first and second
laws of thermodynamics take the form: $ dE = dQ - dW = TdS - PdV.$ Both energy $E$
and entropy $S$ are equilibrium ``state functions'' independent of the path taken
to reach that state.

Among the concepts connecting the atomistic and thermodynamic descriptions entropy
is relatively exotic. Two usual pictures representing entropy are respectively [1] the
many-dimensional phase-space volume consistent with a system's longtime trajectory,
and [2] the cumulative total of heat divided by temperature on a longtime reversible
path from some agreed upon standard state. Unlike energy, pressure, and temperature
knowing the detailed microscopic state (coordinates and momenta) is not enough to
guess the entropy. The  anthropomorphic nature of entropy requires path-dependent
details, either past history or future predictions. The simplest route to entropy
for a simple fluid begins with Carnot's thought experiment coupling the fluid to an
ideal gas constrained to undergo macroscopic changes in temperature and volume.

Accordingly, we describe Carnot's four-part ideal-gas cycle and
use that thought experiment construction to define and characacterize entropy. This is
a traditional textbook path\cite{b1}. To it we add illustrative microscopic simulations
of such cycles for both the ideal gas and an atomistic dense fluid. The simulations
use both isothermal and adiabatic extensions of Hamiltonian molecular dynamics. For
simplicity we study two-dimensional fluid models, both microscopic and macroscopic,
throughout.

The Carnot Cycle itself describes a continuous reversible path of equilibrium states:
two expansions and two compressions. The example  illustrated in {\bf Figures 1 and 2}
is calculated for a two-dimensional monatomic classical ideal gas with the conventional
thermal and mechanical equations of state :
$$
PV = NkT = E = \sum^N [ \ p_x^2 + p_y^2 \ ]/(2m) \ \
[ \ {\rm Two-Dimensional \ Ideal \ Gas} \ ] \ .
$$
The specific equations describing the four-part cycle\cite{b1} in the figures
shown here connect four sets of $(P,V)$ states:
$$
PV = 1 ; \ PV^2 = 1 ; \ PV = 1/2 ; \ PV^2 = 1/2 \ .
$$
Here the Volume $V$ varies from 1/2 to 2 and back in the following way:
$$
(2,1/2)\stackrel{PV = 1    }{---\rightarrow}
(1,1)  \stackrel{PV^2 = 1  }{---\rightarrow}
(1/4,2)\stackrel{PV = 1/2  }{---\rightarrow}
(1/2,1)\stackrel{PV^2 = 1/2}{---\rightarrow} (2,1/2) \ .
$$
These two-dimensional illustrations are adequate and ideally suited to the clarification
of mechanical and thermodynamic concepts.

This specific cycle illustrates the state-function nature of entropy $S$ defined
by the integral of the heat transfer, weighted with the inverse temperature:
$\Delta S \equiv \int dQ/T$. The second and fourth adiabatic parts of Carnot's
cycle are free of heat transfer so that we need only consider the first and third in
entropy calculations. For an ideal gas there is no potential energy so that these
isotherms are also isoenergetic:
$$
0 \equiv dE = dQ - dW = TdS - PdV \longrightarrow
$$
$$
 dQ = dW = TdS = PdV = NkT(dV/V) \longrightarrow
$$
$$
(dQ/T) = dS = Nkd\ln V \ .
$$
The power-law nature of ideal-gas thermodynamics shows that the cyclic integral
of $dQ/T$ vanishes for this and any other Carnot Cycle. An arbitrary ideal-gas
Carnot cycle can be subdivided into a grid of infinitesimal isothermal/adiabatic
$dP \times dV$ clockwise cycles. Then all the internal cycle integrals cancel
their neighbors' contributions. The only exceptions are those perimeter integrals
forming the boundary of the larger arbitrary cycle.

Provided that the ideal-gas heat transfers are divided by their temperatures this
same conclusion holds for a general fluid matching the thermal heat transfers of
the ideal gas. The match is as perfect as is the cycle reversible. Thus the
cyclic integrals of $(dQ/T)$ vanish both for the ideal gas and for a general
fluid, showing that the entropy $S$ is a state function for both these materials.

\noindent
\begin{figure}[h]
\includegraphics[width=3 in,angle=-90.]{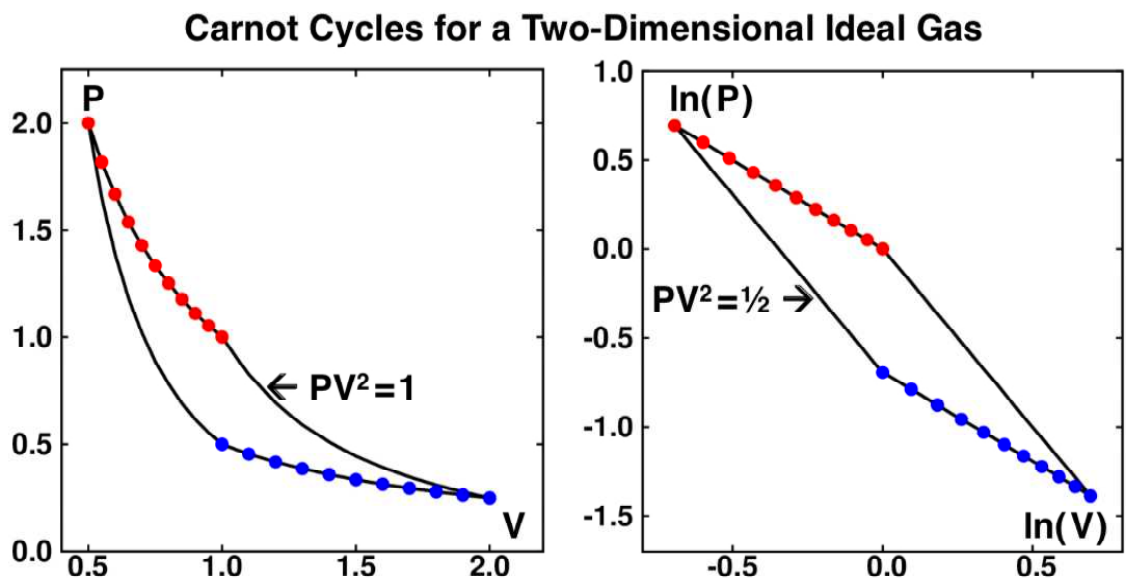}
\caption{
Sadi Carnot's thermodynamic cycle composed (clockwise from top left) of two twofold
expansions, the first isothermal and shown in red, with $T=1$ and $1/2 < V < 1$. The
second expansion, shown black, is adiabatic (no heat transferred) with $1 < V < 2$.
The expansions are followed by two twofold compressions, the first isothermal and shown
in blue with $T=1/2$. The second and last adiabatic compression returns the ideal gas
to the initial state of the cycle, $(P,V,T) = (2,1/2,1)$. The net work done and heat
taken in correspond to the area enclosed by the cycle in the left plot. The right
plot illustrates the powerlaw forms of the ideal-gas mechanical and thermal
equations of state. Notice that the parallel lines show that the isothermal entropy
changes, $\pm \ln(V_{\rm hot}/V_{\rm cold})$ have equal magnitudes so that $\oint dQ/T$
vanishes. As half of the ``hot''-reservoir heat taken in is discharged to the ``cold''
reservoir at the lower temperature of (1/2), the ``efficiency'' of this demonstation
cycle is ``work done''/``heat in'' = 50\%.
}
\end{figure}
\begin{figure}[h]
\includegraphics[width=3 in,angle=-0.]{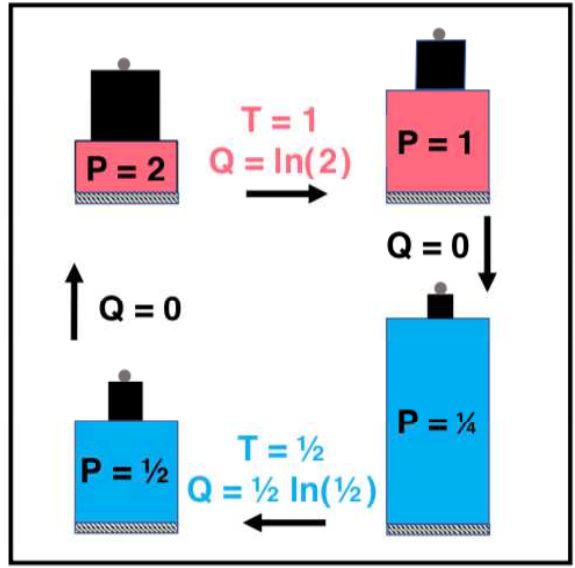}
\caption{
Sadi Carnot's thermodynamic cycle connects two ``hot'' states (at the top, with
temperature 1) and two ``cold'' states (at the bottom, with temperature 1/2).  The
vertical adiabatic processes include no heat transfer so that the two isokinetic
horizontal processes, with $\Delta Q/T = \pm dS$ give a net entropy change of zero,
showing that entropy is a state function.  Coupling the isothermal portions of the
reversed cycle to a general fluid leads to the same conclusion. So long as the
entire cycle is ``reversible'' the net entropy change vanishes: $\oint dQ/T = 0.$
Thus entropy is a state function for a general fluid model.
}
\end{figure}

\pagebreak

\section{Carnot Cycle Simulation with Molecular Dynamics}

In March 1980 Hoover, Ladd, Hickman, and Holian described a Doll's Tensor
Hamiltonian incorporating either a shear or a volumetric strain rate conforming
to the use of periodic boundary conditions\cite{b2}. See the corresponding snapshots
in {\bf Figure 3}. The volume strain shown at the left doubles the area. The
shear strain at the right imposes a displacement in the $x$ direction varying
linearly in $y$. Volumetric strain is present in all four segments of the Carnot
Cycle.\cite{b2} If the four deformations proceed at constant rates notice that
the cycle necessarily includes four points where velocity is discontinuous.

The Hamiltonian for the Carnot Cycle problem, has the following form:
$$
{\cal H}(q,p) = \Phi(q) + K(p) + \dot \epsilon \sum_i (x_ip_{x_i} + y_ip_{y_i}) \ 
[ \ {\rm Doll's \ Tensor} \ ] \ .
$$
Here $\dot \epsilon$ is the macroscopic strain rate,
$$
\dot \epsilon = (du_x/dx) = (du_y/dy) = (1/2)(\dot V/V) \ . 
$$
For simplicity we choose both Boltzmann's constant $k$ and the particle mass $m$
equal to unity throughout. The Doll's Tensor idea is well suited to simulating
Carnot's Cycle, both the adiabatic and the isothermal segments.  Let us detail the
equations of motion.

\subsection{Adiabatic Equations of Motion for Carnot's Cycle}

For each of the $N$ particles in a central $L \times L$ periodic cell centered
on the origin,  $|x| \ {\rm and} \ |y| \ < \ (L/2)$, the Doll's Tensor adiabatic
equations of motion incorporate the volumetric strain rate $2\dot \epsilon$:
$$
\{ \ \dot x_i  = p_{x_i} + \dot \epsilon x_i \ ;
\ \dot p_{x_i} = F_{x_i} - \dot \epsilon p_{x_i} \ ,
   \ \dot y_i  = p_{y_i} + \dot \epsilon y_i \ ;
\ \dot p_{y_i} = F_{y_i} - \dot \epsilon p_{y_i} \ \} \ .
$$

The isotropic pressure $P$ is given by the Virial Theorem:
$$
PV = \sum_i (p_{x_i}^2 + p_{y_i}^2)/2 + \sum_{i < j} (1/2)r_{ij}\cdot F_{ij} \ ,
$$
with
$$
r_{ij} = r_i-r_j \ ; \ F_{ij} = -\nabla_i\phi(|r_i-r_j|) \ .
$$
The first sum above is over particles and the second is over particle pairs.
The Virial Theorem pressure gives the instantaneous rate of energy change:
$$
\dot E = \dot \Phi + \dot K \equiv P\dot V = -2\dot \epsilon PV \ .
$$
These motion equations are ideally suited to the adiabatic deformations in
Carnot's cycle, with $\dot \epsilon$ positive for the two expansions and
negative for the subsequent compressions. The isothermal constraint could be
added by the smooth thermostatting provided by Nos\'e-Hoover mechanics\cite{b3}
or by velocity scaling at the end of each timestep:
$$
\{ \ \dot p_{x_i} = F_{x_i} - \zeta \ p_{x_i} \ ;
   \ \dot p_{y_i} = F_{y_i} - \zeta \ p_{y_i} \ ;
   \ \dot \zeta = [(K/N) - T]/\tau^2 \ \} \ . 
$$
Trials, with orders of magnitude variation in the Nos\'e-Hoover relaxation time
$\tau$, weren't particularly ``useful'', in terms of reducing the fluctuations
in pressure and temperature in the course of the cycle. Instead, we successfully
used both``Gaussian Mechanics''\cite{b4} and velocity rescaling. Both these
algorithms make temperature a constant of the motion. With that choice a 576-particle
system with 2304 motion equations is a convenient size for numerical work. We discuss
computational applications of the underlying motion equations in the following
Sections.

\begin{figure}[h]
\includegraphics[width=4 in,angle=-0.]{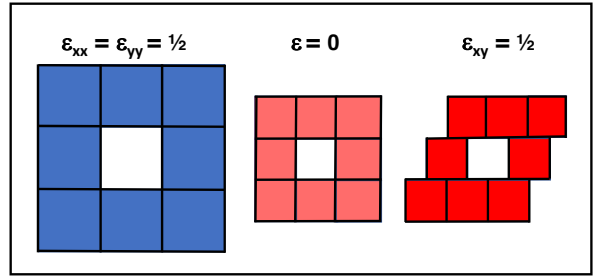}
\caption{
The central square, with $|x|<L/2$ and $|y|<L/2$ is shown in white. The 
initial configuration, in the middle, is surrounded by eight periodic
images. Expansion, which typically leads to cooling, is indicated at the
left and shear, which leads to minor heating, is shown to the right.
These boundary conditions lead naturally to the nearest-image convention
for computing the forces:
{\tt if(xij < -L/2) xij = xij + L ; if(xij > +L/2) xij = xij - L.}
The Carnot cycle of {\bf Figure 1} is composed entirely of volumetric strains,
with $\epsilon_{xx} = \epsilon_{yy}$.
}
\end{figure}

\subsection{Isothermal Equations of Motion for Carnot's Cycle}

Adding instantaneous control of the kinetic energy $\sum (p^2_x + p^2_y)/2$
augments the adiabatic motion equations to include the friction coefficient
$\zeta$:
$$
\{ \ \dot x_i  = p_{x_i} + \dot \epsilon x_i \ ;
\ \dot p_{x_i} = F_{x_i} - \dot \epsilon p_{x_i} - \zeta p_x \ ,
   \ \dot y_i  = p_{y_i} + \dot \epsilon y_i \ ;
\ \dot p_{y_i} = F_{y_i} - \dot \epsilon p_{y_i} - \zeta p_y \ \} \ .
$$
Multiplying the motion equations to compute the kinetic energy $K$
constraint, $\sum (p_x \dot p_x +  p_y \dot p_y) \equiv 0$ provides an
explicit equation for the friction coefficient and the resultant motion
equations for the isothermal segments of the cycle:

$$
\zeta = \sum_i F_i\cdot p_i/(2K) \ ; \{ \ \dot q = p + \dot \epsilon q \ ; \
\dot p = F - \zeta p \ \} \ .
$$

Numerical exploration of the adiabatic and isothermal motion equations shows
that both approaches are well-behaved, stable, and relatively close to our
expectations based on the thought-experiment version of the Carnot cycle. A
useful modification of the isothermal motion equations instead adds a
velocity rescaling operation, $p_i \rightarrow p_i \times \sqrt{2NT/\sum p^2}$,
at the conclusion of each adiabatic timestep. Accumulating the kinetic energy
changes due to these rescalings provides the computational version of heat
gain or loss $dQ$. These energy increments are necessary in evaluating the efficiency
and the dissipation associated with the cycles.  We describe our computational
experience with the Carnot-Cycle molecular dynamics next.

\pagebreak

\section{Computational Carnot Cycles for a Knudsen Gas}

A practical approach to the nonequilibrium molecular dynamics of the Carnot
Cycle is to begin with a conventional four-body computer program incorporating static
periodic boundary conditions. Once that isoenergetic programming is successful
it is straightforward to introduce dynamic boundaries and to extend Knudsen-gas and
dense-fluid codes to larger systems for which fluctuations are smaller. We chose
$N=24^2 = 576$ to illustrate the present work, large enough that the fluctuations are
small, but with the system still small enough that laptop problems can be completed in a
few hours' time. Equilibrium simulations with zero strain rate suggested a fourth-order 
Runge-Kutta timestep $dt = 0.001$ with kinetic temperatures of 0.5 and 1.0 for the
``cold'' and ``hot'' isothermal segments of the cycle. The minimum and maximum
densities imposed on the cycles were $(1/2)$ and $(2)$, matching our 1991 textbook
example problem\cite{b1}.

{\bf Figure 4} shows two views of a (collisionless) Knudsen-Gas cycle. The
lack of collisions would seem to suggest, wrongly, that the initial velocity
distribution remains unchanged.  The adiabatic coordinate changes are paired
with momentum changes keeping the phase volume $dqdp$ fixed. The twofold
density increases and decreases correspond to twofold increases and decreases
in temperature. In the isothermal segments the coordinates expand or contract
but the momenta are unchanged. For isothermal processes the kinetic temperature
$\langle p_x^2 \rangle = \langle p_y^2 \rangle$ is constant, imposed with the
reversible friction $\zeta$ or by velocity rescaling. The discontinuous velocities
when the four segments begin and end cause no particular computational difficulties.

For the adiabatic segments the Knudsen gas equations of motion,
$$
\{ \ \dot q = p + \dot \epsilon q \ ; \ \dot p = - \dot \epsilon p \ \} \ ,
$$
can be integrated analytically (though Runge-Kutta integration is both
faster and simpler). Each of the Knudsen-gas momenta varies exponentially
in time in the adiabatic segments:
$$
p(t) = p(0)e^{-\dot \epsilon t} = p(0)e^{-\epsilon} \ ,
$$
where $\epsilon$ is the strain and $\dot \epsilon$ is the strain rate, for
simplicity chosen constant for each segment of the cycle.  In the second
segment of the cycle, the adiabatic expansion phase, the volume is doubled
and the final one-dimensional strain is $\sqrt{2}$. During the expansion
the $x$ and $y$ components of each velocity are reduced by the same factor.
The initial Gaussian, with kinetic temperature $\langle p_x^2 \rangle =
\langle p_y^2 \rangle = 1$, is reduced in amplitude. This results in the final
cooler kinetic temperature $\langle p_x^2 \rangle = \langle p_y^2 \rangle =
(1/2)$. The collisionless molecular dynamics reproduces the thought-experiment
cycle of {\bf Figure 1} nearly perfectly, as is shown in {\bf Figure 4}.

Analysis of the cycle shows that half the heat taken in at $T=1$ is given
to the cold reservoir at $T = (1/2)$ with the other half converted to work.
Each clockwise thermodynamic cycle performs net work equal to $\ln (\sqrt{2})$
per particle.

\begin{figure}[h]
\includegraphics[width = 2 in,angle=-90.]{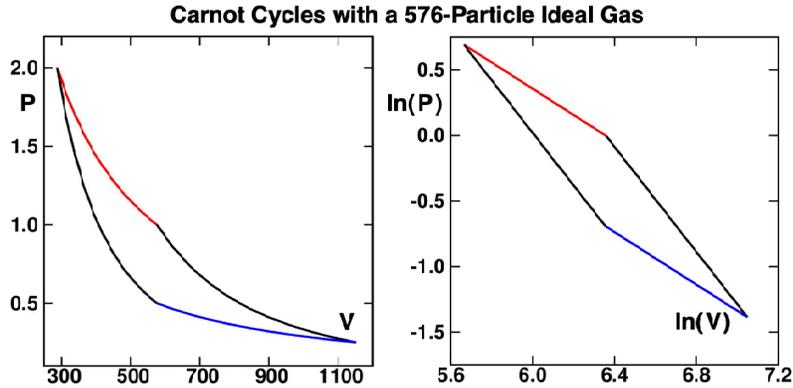}
\caption{
Knudsen Gas Carnot Cycle with 576 ideal-gas particles. The initial
distribution of momenta, chosen Gaussian, remains Gaussian throughout
as the temperature varies from 0.5 to 1.0. The four segments of the
cycle all correspond to twofold changes in density. The heat taken in
at $T = 1$, $\int_{1/2}^1d\ln V$, $\ln (2)$ per particle, is twice that
given off at $T=(1/2)$, with the difference equal to the work done,
$288\times \ln (2)/2 = 199.6$, corresponding with the thought experiment
of the theoretical cycle of {\bf Figure 1}.
}
\end{figure}

\pagebreak

\section{Simulating Carnot Cycles for a Soft-Disk Dense Fluid}

For a ``realistic'' dense-fluid model we use a simple purely-repulsive pair
potential with a finite range of unity:
$
\phi(r<1) = (10/\pi)(1-r)^3 \longrightarrow \int_0^1 2\pi r\phi(r) \equiv 1 \ .
$
Steady-shear simulations with a similar short-ranged repulsive potential,
$100(1-r^2)^4$ suggest that size and strain rate dependences act to reduce
the magnitude of the shear viscosity by no more than a few percent\cite{b5}.
Because Carnot-cycle deformations are volumetric, at constant shape, dissipation
occurs in the form of ``bulk viscosity'' rather than shear.  In 1971 David Gass
estimated the bulk viscosity for hard disks\cite{b6}. Setting the soft-disk
potential equal to a kinetic temperature of unity gives an effective diameter
of $1 - \sqrt[3]{(\pi/10)} = 0.3202$ from which Gass' bulk viscosity is roughly
0.04.

We choose the same temperatures, 1 and (1/2), and the same density range,
$(1/2) < \rho = (N/V) < 2$ as in the Knudsen Gas example. {\bf Figure 5} shows
the resulting cycles for times of 1, 2, 4, and 8 for each of the four segments.
Segment times of 256 or 512 are reasonable problems taking only a few hours
of laptop time. The longer slower problems show much smaller pressure
fluctuations than the faster problems of {\bf Figure 5}. The initial value,
$PV/N$ near 1.8, necessarily exceeds the ideal-gas value of unity. Likewise,
the minimum, around 0.6, exceeds the ideal-gas value $T = (1/2)$.

The larger slower problems are suitable for thermodynamic analyses of
fluctuations.  The isokinetic friction coefficient, equivalent to the
velocity rescaling factor:
$$
\zeta = \frac{\sum_i F_i\cdot p_i}{\sum_i p_i^2}                                           
$$
describes the entropy change in thermostatted flows and exhibits large
fluctuations in small systems with rapid deformation. We experimented
with dozens of combinations of strain rate, timestep, system size and
numbers of cycles. A useful estimate of finite-system effects can be
based on the series of strain rates $\dot \epsilon = (1/2)^n$. Hydrodynamics
suggests an entropy production varying as the square of the strain rate,
so that for a fixed number of cycles the efficiency should show a loss
proportional to the strain rate.

For the series $n = 1 \ {\rm to} \ 5$ we found efficiencies increasing from 88\% to 98\%
of the ideal. In the latter case the lost work per cycle, about 5, should
be of order $32(2/32)^2 N \eta_V$, corresponding to a bulk viscosity of
order 0.05, close to the hard-disk estimate based on Gass' work. In all,
the agreement of the simulations with expectations is quite satisfactory.
The unaesthetic nature of the velocity discontinuities can only be avoided
by adding complexity to the analysis, already clouded by the relatively
wide difference in the strain rates.

\begin{figure}[h]
\includegraphics[width=2.5 in,angle=-90.]{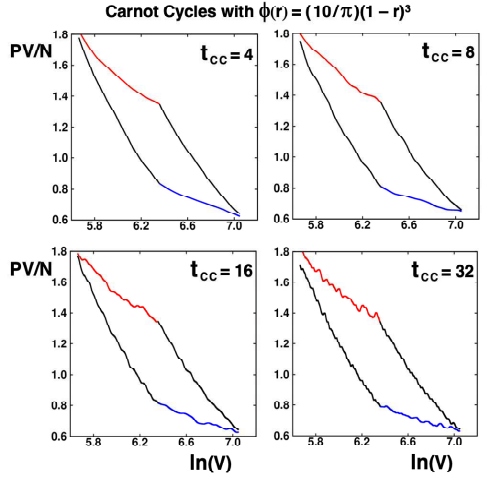}
\caption{
  Soft-disk Carnot Cycle with 576 particles. The pair potential is
$\phi(r) = (10/\pi)(1-r)^3$. In the cycle $0.5 < T < 1$ and 
$(1/2) < \rho < 2$. The work per particle done in the cycle
is the enclosed area, roughly 0.4. The total work, larger by a factor of
$N = 576$, increases from about 200 to 234 as the cycle time $t_{CC}$ is
increased from 4 to 128. The efficiency of the slowest cycle, with $t_{CC}
= 512$, is 98\% of the ideal 50\%.
}
\end{figure}

\begin{figure}[h]
\includegraphics[width=1.4 in,angle=-90.]{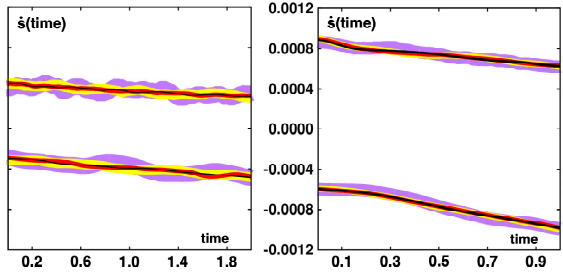}
\caption{
Entropy gained and lost, by ``hot'' and ``cold'' reservoirs respectively
in following the dense-fluid Carnot Cycle with periods of 4 (on the right)
and 8 (on the left). The numbers of cycles in the averaging vary from
50 to 1000 (left) or 100 to 2000 (right) with the shorter runs (purple)
exhibiting larger fluctuations than the longer ones (black). Precise
averages require on the order of 100 cycles for 576 particles. The entropy
lost by the hot reservoir exceeds that gained by the cold one. The
efficiencies of the cycles vary from 88\% of the reversible ideal, for the shortest run,
to 98\% for the longest. Simulations with $dt = 0.01$ are consistent
with the $dt=0.001$ used here.
}
\end{figure}

\pagebreak

\section{Summary}

In Gibbs' and Boltzmann's microscopic statistical mechanics entropy is related to
the longtime occupation of states in phase space, $S(N,E,V) = k\ln \Omega$.
In macroscopic thermodynamics entropy is related to the longtime integrated,
slow and reversible, uptake of heat divided by temperature,
$S(N,E,V) = \int_{-\infty}^t dQ/T$.  Here we have explored these two very different
views of entropy for two material models, a Knudsen gas, with a collisionless
Maxwell-Boltzmann velocity distribution, and a simple fluid, with a short-ranged
purely-repulsive atomistic pair force. We have introduced and used two
different versions of nonequilibrium molecular dynamics to implement Carnot Cycles
for the two microscopic models.

The Carnot Cycle is fundamental to the connection of phase volume $\Omega$ to
integrated heat $Q$ and the kinetic temperature $T$. The cyclic conversion of heat
to work cleanly separates macroscopic isothermal and adiabatic volume changes in an
elegant way applicable to general fluid models. This cyclic connection of ideal-gas
states to fluid states through work and heat provides the simplest possible
illustration connecting the microscopic and macroscopic definitions of entropy.

The Carnot Cycle is perfectly suited to two relatively new versions of mechanics:
[1] an isothermal mechanics based on Gauss' Principle of Least Constraint, where the
kinetic temperature is a constrained variable; and [2] an adiabatic mechanics, based
on the Doll's-Tensor Hamiltonian,
$$
{\cal H}_{\rm Doll's}(q,p) = {\cal H}_{\rm Equilibrium}(q,p) +
\dot \epsilon \sum_i q_ip_i \rightarrow \dot E = -P\dot V \ .
$$
This Hamiltonian introduces adiabatic time-dependence into the conservation of
mechanical energy, $dE/dt = -PdV/dt = -2\dot \epsilon PV$ through the macroscopic
strain rate $\dot \epsilon$.

Our implementation of the ideal-gas and simple-fluid cycles illustrates the
connection of atomistic dynamics to macroscopic thermodynamics. We characterized
the dissipative heat for a series of strain rates and related it to bulk
viscosity estimates from kinetic theory.

The Carnot cycle is an ideal teaching tool for relating thermodynamics to
statistical mechanics. The models suggest further research opportunities. In
particular it is desirable to formulate cycles with continuous strainrates
and to further detail the microscopic bulk-viscosity mechanism underlying these
illustrations of the Second Law of Thermodynamics.

\pagebreak

\section{Acknowledgment}
We thank Karl Travis and Ed Smith for their cogent comments and suggestions.

\end{document}